\documentclass[aps,twocolumn]{revtex4}
\usepackage{graphics}
\usepackage{epsfig}
\usepackage{amsmath}

\newcommand{\Ket}[1]{|#1\rangle}

\newcommand{\BraKet}[2]{\langle#1|#2\rangle}
\newcommand{\sqrtsw}{$\sqrt{\text{SWAP}}$\ }
\newcommand{\ucnot}{U_{\text{CNOT}}}
\newcommand{\uphase}{U_{\text{PHASE}}}
\newcommand{\uswap}{U_{\text{SWAP}}}
\newcommand{\usqrtsw}{U_{\sqrt{\text{SWAP}}}}
\newcommand{\bc}[1]{b^{\dagger}_{#1}}
\newcommand{\ba}[1]{b_{#1}}
\newcommand{\bcs}[1]{{b'}^{\dagger}_{#1}}
\newcommand{\bas}[1]{{b'}_{#1}}
\newcommand{\tr}{\text{tr}}
\newcommand{\0}{\Omega}

\begin{document}
\title{{\bf Quantum computing in optical microtraps }\\
{\bf based on the motional states of neutral atoms}}
\author{K.~Eckert$^{a}$, J.~Mompart$^{a,b}$, X.~X.~Yi$^{a,c}$,
 J.~Schliemann$^{d}$, D.~Bru{\ss}$^{a}$, G. Birkl$^{e}$, and M. Lewenstein$^{a}$}
\affiliation{
\smallskip
$^a$Institute of Theoretical Physics, University of Hannover,
Appelstr. 2, D-30167, Hannover, Germany\\
$^b$ Departament de F\'{\i}sica, Universitat Aut\`onoma de Barcelona, 
E-08193, Bellaterra, Spain \\
$^c$ Institute of Theoretical Physics, Northeast Normal University, 
Changchun 130024, China \\
$^d$Department of Physics and Astronomy, University of Basel, CH-4056
Basel, Switzerland\\
$^e$ Institute of Quantum Optics, University of Hannover, Welfengarten 1, D-30167, Hannover, Germany}
\date{\today}
\begin{abstract}
We investigate quantum computation with neutral atoms in optical microtraps
where the qubit is implemented in the motional states of the atoms, i.e., in
the two lowest vibrational states of each trap. The quantum gate operation is
performed by adiabatically approaching two traps and allowing tunneling and
cold collisions to take place. We demonstrate the capability of this
scheme to realize a square-root of swap gate, and address the problem of double
occupation and excitation to other unwanted states. We expand the
two-particle wavefunction in an orthonormal basis and analyze quantum correlations
throughout the whole gate process. Fidelity of the gate operation is
evaluated as a function of the degree of adiabaticity in moving the traps.
Simulations are based on rubidium atoms in state-of-the-art optical
microtraps with quantum gate realizations in the few tens of milliseconds
duration range. 

\end{abstract}

\maketitle

\section{Introduction}

The development of tools to prepare, manipulate and measure the quantum
state of a physical system represents one of the great challenges of modern
science and, in particular, it is essential for applications in quantum
information processing such as quantum computing. At present a few systems have been
identified that should permit quantum computation: molecules in the
context of NMR \cite{NMR}, ion-traps \cite{ion}, cavity QED with photons and atoms \cite{QED},
solid state devices such as quantum dots \cite{swap,dots1,dots2},
and trapped neutral atoms \cite{na1,na2,na3,na4,na5}. For Rydberg atoms in
high Q cavities the engineering of entangled states and the implementation
of quantum logic have been demonstrated \cite{qedExp}, furthermore a quantum
gate has been performed between the internal and external degrees of freedom of
an ion in a trap \cite{ionExp}. In NMR systems, quantum algorithms on a few qubits, e.g.,
Shor's factoring algorithm, have been reported \cite{NMRexp}.

Neutral atoms are promising candidates for quantum computing for at least two
reasons: (i) techniques of cooling and trapping atoms are by now very
well established \cite{BEC}; and (ii) they are comparatively less sensitive
to decoherence, e.g., interaction with the `classical' environment. Neutral
atoms can be stored and manipulated in optical lattices \cite{latt}, 
standard dipole traps \cite{single}, and
microtraps \cite{mag,opt1,opt2,opt3}. In particular, magnetic \cite{mag} and
optical microtraps \cite{opt1,opt2,opt3} offer an interesting perspective
for storing and manipulating arrays of atoms with the eventual possibility
to scale, parallelize, and miniaturize the atom optics devices needed in
quantum information processing. Moreover, optical microtraps can take
advantage of the fact that most of the current techniques used in atom
optics and laser cooling are based on the optical manipulation of atoms \cite
{opt1}.
Many of the requirements for the implementation of quantum computation
\cite{divi} have been recently demonstrated in optical microtraps containing 
$\sim 100$ atoms per site \cite{opt3},
e.g., selective addressing of single trap sites, and initializing 
and reading-out of quantum states in each site. 
In addition, the possibility to store and detect single atoms in optical dipole traps 
has been reported \cite{single}. 

With the demonstration of single-qubit gates being straight forward, what remains 
to be experimentally demonstrated is the capability of these optical microtraps 
to perform two-qubit quantum gates.
The most prominent examples of such gates include the CNOT gate, the phase gate
 and the \sqrtsw gate \cite{swap,CNOT}. The latter transforms states
$\Ket{0}\Ket{1}$ and $\Ket{1}\Ket{0}$, written in the computational basis, to
maximally entangled states, while leaving $\Ket{0}\Ket{0}$ and $\Ket{1}\Ket{1}$
unaffected, in such a way that after the successive application of two \sqrtsw gates 
the states of the qubits are interchanged.
Each one of these two-qubit gates, together with arbitrary single qubit operations,
is universal, i.e., allows to perform any quantum algorithm. 
In practice, the particular two-qubit gate to be implemented will depend on
the physical system under consideration.

With respect to neutral atoms, several different physical mechanisms 
to perform two-qubit gates have been proposed, ranging from cold controlled collisions 
\cite{na1,na2} and dipole-dipole interactions \cite{na3,na4,na5} to purely geometric
quantum evolution \cite{holonomic}.
In the cold collisional case, a two-qubit
phase gate was proposed by adiabatically approaching two traps \cite{na1}
or by instantaneous state-selective switching of the trapping
potentials \cite{na2}. In both cases the qubit was encoded in some internal degrees of
freedom of the atoms, e.g., spin, Zeeman or hyperfine levels. For cold collisions
to take place the atoms have to be brought to close distances,
such that their quantum statistical nature has to be taken into account.
A detailed study of the role of the bosonic or fermionic character of particles
in the context of quantum information in atomic waveguide structures
has been done by E.~Andersson {\it et al.} \cite{Andersson1,Andersson2}.

Here we address the problem of implementing a quantum gate by adiabatically
approaching two bosonic atoms, each stored in a different microtrap.  
In contrast to the proposals mentioned above, we assume the qubit to be
implemented in the motional states of the atoms, i.e.\ an atom in
the ground or the first vibrational state of the trap represents 
$|0\rangle $ or $\Ket{1}$, respectively.
Note that, as for the ion-trap
case, the observation of neutral atoms cooled down 
to the ground and first vibrational states as
well as superposition states in one-dimensional traps has been achieved \cite{vib}.
To perform the gate operation, we apply the steps
outlined in Fig.~1. Initially, the two microtraps are far apart such that
the interaction between the two atoms is negligible.
Then we adiabatically move both traps close together such that
tunneling and cold controlled collisions become important. The dynamics of
this process strongly depend on the particular motional state of the atoms
and we can make use of this fact to control the interaction
such that, after the eventual separation of the traps, the desired gate operation 
is realized with each trap again containing only one atom.

To be more specific we consider here laser-cooled rubidium atoms stored in optical microtraps
\cite{opt3}, assuming that each trap contains initially only one atom.
We will show that the {\it \sqrtsw gate} is the most natural
quantum gate to be implemented when the qubit is encoded in the motional states of the atoms
and interaction takes place through tunneling and cold collisions.
This result applies to both, $^{85}$Rb and $^{87}$Rb, although they
have negative and positive scattering length, respectively.
In particular, we will demonstrate that a quantum gate of $\sim 20$ milliseconds
duration can be performed in state-of-the-art optical microtraps. 
Very recently, E.~Charron {\it et al.}~\cite{williams} have
proposed the implementation of a {\it phase gate} in an optical lattice
with, as it is done also here, the qubit encoded in the motional states. 
In this case, a controlled interference set-up was proposed to
perform a high-fidelity gate with operation time of $38$~ms.

The paper is organized as follows. In Section {\rm II}, we introduce the
physical model. Section {\rm III} is devoted to the implementation of the
\sqrtsw gate. In Section IV\ we discuss some practical
considerations. And, finally, section {\rm V} summarizes the results and
presents the conclusions.

\section{Model}

In this section we will first write down the Hamiltonian for the two atoms
stored in the microtraps and discuss the interaction mechanism. We will
introduce a time-dependent orthonormal set of single-particle states for each
trap that is also orthogonal to the states of the other trap for arbitrary
distances between the two traps. These single-particle states will make it possible
to expand the wavefunction in a set of two-particle orthonormal states. This
representation has two important advantages: (i) it allows to compute entanglement
throughout the whole gate process; and (ii) it strongly reduces the computational
time required to simulate a quantum gate operation with respect to a direct
numerical integration of the Schr\"{o}dinger equation for the two-particle
spatial wavefunction. Finally, we will discuss the physical implementation
of the qubits and its implications for the quantum gate operations.

\subsection{Hamiltonian}

The Hamiltonian governing the dynamics of the two atoms in a time-varying
particle-independent trapping potential $V(\vec{r},t)$ can be written as 
\begin{equation}
H=\sum_{i=1,2}\left[ \frac{\vec{p}_{i}^{\,2}}{2m}+V(\vec{r}_{i},t)\right]
+U\left( \vec{r}_{1}-\vec{r}_{2}\right) ,
\end{equation}
where $m$ is the mass of the atoms, $\vec{r}_{i}$ and $\vec{p}_{i}$ are the
(three-dimensional) position and momentum operators for atoms $1$ and $2$,
and $U\left( \vec{r}_{1}-\vec{r}_{2}\right) $ accounts for the interaction
between the two atoms.

To simplify the problem, we take the trapping potential shape to be
time-independent along $y$ and $z$ directions: 
\begin{equation}
V(\vec{r},t)=v(x,t)+v_{p}(y)+v_{p}(z),
\end{equation}
and assume much stronger confinement in $y$ and $z$ directions than in $x$,
such that transverse excitations can be neglected. In fact, we will consider
that both atoms are cooled down to the $y$ and $z$ vibrational ground states
and remain there during all the interaction process. Explicitly, we take the
following one-dimensional potential to describe the two microtraps separated
by a distance $2a(t)$: 
\begin{equation}
v(x,t)=\frac{m\omega _{x}^{2}}{2}\left[ \left( x+a(t)\right) ^{2}\theta
(-x)+\left( x-a(t)\right) ^{2}\theta (x)\right]  \label{trapx}
\end{equation}
where $\omega _{x}$ is the trapping frequency in the $x$ direction, 
and $\theta (x)$ is the step function.

The temporary variation of the trap distance is sketched in Fig.~1.
Initially the traps are separated by a distance $2a_{\text{max}}$. The process
of slowly approaching them to a minimum separation $2a_{\text{min}}$ takes a time
$t_r$ and is modeled by the first quarter of a period of a cosine.
Then we let the atoms interact for a time $t_i$ and, finally,
we slowly separate the traps.

\begin{figure}
%
%
\begin{center}
\includegraphics[width=0.8\linewidth]{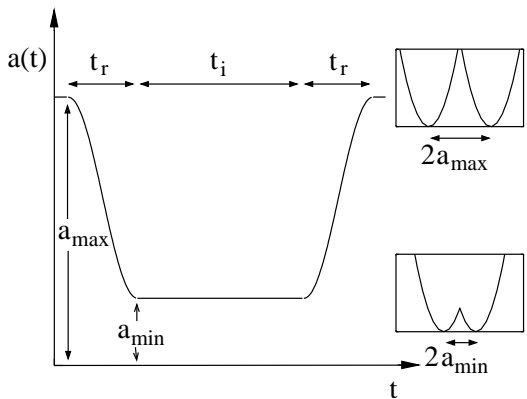} 
\end{center}
\label{time}
\caption{Separation of the traps as a function of time. 
$t_r$ and $t_i$ are the approaching/separating and interaction times, respectively.
At $a_{\max}$ atoms located in different traps do not interact  
while at $a_{\min}$ tunneling and cold collisions take place.  
}
\end{figure}

For cold bosonic atoms, the dominant collisional interaction is due to $s$-wave
scattering \cite{na2}, which can be described by a contact
potential of the form 
\begin{equation}
U(\vec{r}_{1}-\vec{r}_{2})=\frac{4\pi a_t\hbar ^{2}}{m}\delta ^{3}(\vec{r}%
_{1}-\vec{r}_{2}),
\end{equation}
where $a_t$ is the $s$-wave scattering length of the rubidium atoms,
e.g., in the spin triplet $a_t=-369$~$a_0$ for $^{85}$%
Rb and $a_t=106$~$a_0$ for $^{87}$Rb with $a_0$ being the Bohr radius. 
As long as both atoms remain in the transverse
vibrational ground states, we can integrate out the corresponding degrees of
freedom and obtain an effective one-dimensional interaction potential \cite
{na2} 
\begin{equation}
u(x_{1}-x_{2})=2 a_t\hbar \omega _{p}\delta (x_{1}-x_{2}),  \label{scat}
\end{equation}
where $\omega _{p}$ is the transverse trapping frequency. Eqs.~(\ref{trapx})
and (\ref{scat}) allow us to reduce the complexity of the problem to one
dimension.

\subsection{Single-particle states}

We will implement the qubits into the ground and first excited vibrational
states of each trap, i.e., we will use the motional states of the atoms.
When the two traps are far apart, i.e., $a\alpha \gg 1$ with 
$1/\alpha \equiv \sqrt{\hbar/m\omega _{x} }$ being the position uncertainty of the
ground state,
these states are the energy eigenstates of two displaced one-dimensional
harmonic oscillators:
\begin{subequations}
\begin{eqnarray}
\langle x|0\rangle _{L,R} &=&\frac{\sqrt{\alpha }}{\pi ^{1/4}}e^{-\frac{1}{2}%
\alpha ^{2}(x\pm a)^{2}}, \\
\langle x|1\rangle _{L,R} &=&\frac{\sqrt{2\alpha }}{\pi ^{1/4}}e^{-\frac{1}{2%
}\alpha ^{2}(x\pm a)^{2}}\alpha (x\pm a),
\end{eqnarray}
\end{subequations}
with $L$ and $R$ labeling the left and right trap, respectively.
As we approach the two traps, these single-particle states overlap
and are no longer orthogonal. To numerically integrate the 
Schr\"odinger equation and to compute entanglement throughout the
gate process, we construct an orthonormal single-particle basis
for arbitrary distances of the two traps by applying the Gram-Schmidt
method (see Appendix A). If we denote these new single-particle
states by $|\overline{i}\rangle_{s}$ with $i=0,1,2,3...$ and $s=L,R$
then it holds ${}_{s}\BraKet{\overline{i}}{\overline{j}}_{t}=
\delta_{ij}\delta_{st}$. The four states that for large distances correspond 
to the two lowest states of each trap read
\begin{subequations}
\begin{eqnarray}
\langle x|\overline{0}\rangle _{L,R} &=&\langle x|0\rangle _{L,R}\frac{\xi
_{0}^{+}+\xi _{0}^{-}}{2}+\langle x|0\rangle _{R,L}\frac{\xi _{0}^{+}-\xi
_{0}^{-}}{2} \\
\langle x|\overline{1}\rangle _{L,R} &=&(\langle x|1\rangle _{L,R}-\frac{%
xe^{-a^{2}\alpha ^{2}}}{\pi ^{1/4}\alpha ^{3/2}}\langle x|0\rangle _{R,L})%
\frac{\xi _{1}^{+}+\xi _{1}^{-}}{2}  \nonumber \\
\lefteqn{+(\langle x|1\rangle _{R,L}-\frac{xe^{-a^{2}\alpha ^{2}}}{\pi ^{1/4}\alpha
^{3/2}}\langle x|0\rangle _{L,R})\frac{\xi _{1}^{-}-\xi _{1}^{+}}{2}}
\end{eqnarray}
where $\xi _{0}^{\pm }(a)$ and $\xi _{1}^{\pm }(a)$ are given in
Eqs.~(A5).
For large separation of the traps, i.e.\ $a \alpha \gg 1$, we have 
$\xi_i^+=\xi_i^-$ for all $i$ and thus the $\Ket{\overline{i}}_{L,R}$
become the  eigenstates of a single harmonic trap centered at $\mp a$.
Notice that the $\Ket{\overline{i}}_{L,R}$ states have the following symmetry
under parity transformation:
$\langle x|\overline{i}\rangle _{L,R}\mapsto \left( -1\right)
^{i}\langle x|\overline{i}\rangle_{R,L}$. The general proof is given in Appendix A. 
This property obviously holds 
for the $\Ket{i}_{L,R}$, and the $\Ket{\overline{i}}_{L,R}$ are constructed 
such that this symmetry is maintained.

Although above we have written only four states, for all
simulations using these orthogonalized states we will include
all states up to $\Ket{\overline{3}}_{L,R}$. 

\subsection{Two-particle states}

Let us motivate the two-particle basis which we will use. On the one hand, it
must satisfy bosonic statistics, i.e.\ the basis states have to
be symmetric under the permutation of the particles. On the other
hand the Hamiltonian of this system is symmetric with respect to parity
transformation, i.e.\ $H(x)=H(-x)$, and therefore does not couple states
of opposite parity. For this reason we will introduce basis states with well-defined
parity. If for this description we limit ourselves again to the four lowest
single-particle states, then the bosonic two-particle sector forms a ten-dimensional 
Hilbert space. Here, we use the following notation 
$\Ket{\overline{m}(1)}_{s}\otimes\Ket{\overline{n}(2)}_{t}
\equiv\Ket{\overline{m}}_{s}\Ket{\overline{n}}_{t}$ with 1 and 2 labeling the atoms
and $s,t=L,R$. Thus,
the bosonic two-particle basis reads: 
\end{subequations}
\begin{subequations}
\label{01s}
\begin{eqnarray}
|00\rangle ^{+} &=&\frac{1}{\sqrt{2}}\left( \left| \overline{0}\right\rangle
_{L}\left| \overline{0}\right\rangle _{R}+\left| \overline{0}\right\rangle
_{R}\left| \overline{0}\right\rangle _{L}\right) , \\
|01\rangle ^{+} &=&\frac{1}{2}\left( \left| \overline{0}\right\rangle
_{L}\left| \overline{1}\right\rangle _{R}+\left| \overline{1}\right\rangle
_{R}\left| \overline{0}\right\rangle _{L}\right.  \nonumber \\
&&\left. -\left| \overline{0}\right\rangle _{R}\left| \overline{1}%
\right\rangle _{L}-\left| \overline{1}\right\rangle _{L}\left| \overline{0}%
\right\rangle _{R}\right) , \\
|11\rangle ^{+} &=&\frac{1}{\sqrt{2}}\left( \left| \overline{1}\right\rangle
_{L}\left| \overline{1}\right\rangle _{R}+\left| \overline{1}\right\rangle
_{R}\left| \overline{1}\right\rangle _{L}\right) , \\
|\widetilde{00}\rangle ^{+} &=&\frac{1}{\sqrt{2}}\left( \left| \overline{0}%
\right\rangle _{L}\left| \overline{0}\right\rangle _{L}+\left| \overline{0}%
\right\rangle _{R}\left| \overline{0}\right\rangle _{R}\right) , \\
|\widetilde{01}\rangle ^{+} &=&\frac{1}{2}\left( \left| \overline{0}%
\right\rangle _{L}\left| \overline{1}\right\rangle _{L}+\left| \overline{1}%
\right\rangle _{L}\left| \overline{0}\right\rangle _{L}\right.  \nonumber \\
&&\left. -\left| \overline{0}\right\rangle _{R}\left| \overline{1}%
\right\rangle _{R}-\left| \overline{1}\right\rangle _{R}\left| \overline{0}%
\right\rangle _{R}\right) , \\
|\widetilde{11}\rangle ^{+} &=&\frac{1}{\sqrt{2}}(|\bar{1}\rangle _{L}|\bar{1%
}\rangle _{L}+|\bar{1}\rangle _{R}|\bar{1}\rangle _{R}),
\end{eqnarray}
and 
\end{subequations}
\begin{subequations}
\label{01as}
\begin{eqnarray}
|01\rangle ^{-} &=&\frac{1}{2}\left( \left| \overline{0}\right\rangle
_{L}\left| \overline{1}\right\rangle _{R}+\left| \overline{1}\right\rangle
_{R}\left| \overline{0}\right\rangle _{L}\right.  \nonumber \\
&&\left. +\left| \overline{0}\right\rangle _{R}\left| \overline{1}%
\right\rangle _{L}+\left| \overline{1}\right\rangle _{L}\left| \overline{0}%
\right\rangle _{R}\right) , \\
|\widetilde{00}\rangle ^{-} &=&\frac{1}{\sqrt{2}}\left( \left| \overline{0}%
\right\rangle _{L}\left| \overline{0}\right\rangle _{L}-\left| \overline{0}%
\right\rangle _{R}\left| \overline{0}\right\rangle _{R}\right) , \\
|\widetilde{01}\rangle ^{-} &=&\frac{1}{2}\left( \left| \overline{0}%
\right\rangle _{L}\left| \overline{1}\right\rangle _{L}+\left| \overline{1}%
\right\rangle _{L}\left| \overline{0}\right\rangle _{L}\right.  \nonumber \\
&&\left. +\left| \overline{0}\right\rangle _{R}\left| \overline{1}%
\right\rangle _{R}+\left| \overline{1}\right\rangle _{R}\left| \overline{0}%
\right\rangle _{R}\right) , \\
|\widetilde{11}\rangle ^{-} &=&\frac{1}{\sqrt{2}}\left( \left| \overline{1}%
\right\rangle _{L}\left| \overline{1}\right\rangle _{L}-\left| \overline{1}%
\right\rangle _{R}\left| \overline{1}\right\rangle _{R}\right) .
\end{eqnarray}
\end{subequations}
The notation at the l.h.s.\ of Eqs.~(\ref{01s}) and (\ref{01as})
means the following: superscripts $+$ or $-$ indicate that the two-particle
state has positive or negative parity, respectively, 
while the tilde accounts for states where, for $a\alpha\gg 1$,
both atoms are in the same trap, i.e., double-occupancy states.
It is easy to check the symmetry of these two-particle states under
the exchange of the atoms by making use of the parity property of states $%
\left| \overline{i}\right\rangle _{L,R}$ discussed after Eqs.~(7).

In addition it is worth to mention that in our simulations we will consider up to eight
single-particle states which gives rise to a bosonic two-particle 
Hilbert space of 36 states (20 states having positive parity from
which 10 correspond to double occupancy; and 16 states having negative
parity with 10 accounting for double occupancy).
Finally note that the fact that we are able to expand the
wave-function into this finite number of two-particle orthogonal states has
also an important advantage with respect to the time needed for the
simulation of a gate operation.
We have checked the accuracy of the restriction of
the simulation to this subspace by comparing the results of the simulations 
to a direct numerical integration of the Schr\"{o}dinger equation for the two-particle
spatial wavefunction which is about four orders of magnitude slower. 

\subsection{Physical implementation}

We start from two well separated traps, each containing one atom.
In this situation we can neglect the bosonic nature of the particles
and forget about the symmetrization \cite{peresbook}. Only then it 
is possible to speak about well-defined qubits and we choose to introduce
labels $A$ and $B$ for the two qubits by labeling the atom found in
the left trap by $A$ and the atom in the right trap by $B$. 

With the two traps far apart, single qubit operations, e.g., a Hadamard gate, 
can be realized by using two laser pulses in a Raman configuration 
focused solely on one of the traps.
The quantum gate operation between two qubits is much more involved. As we
approach the traps, due to tunneling there will be a non-vanishing probability
to find both atoms in the same trap. Thus we can no longer
distinguish the atoms such that bosonic statistics become important
and the qubits are no longer well defined. If, however, we 
approach and separate the traps in such a way that finally there is again
one atom in each of the well separated traps then we can attribute (new) labels
$A$ and $B$ to them in the same way as before.

These considerations suggest
the following mapping of the states of the computational basis into the two-particle
basis states of Eqs.~(\ref{01s}) and (\ref{01as}):
\begin{subequations}
\label{9}
\begin{eqnarray}
\left| 0\right\rangle _{A}\left| 0\right\rangle _{B} &\rightarrow&
|00\rangle ^{+} \\
\left| 0\right\rangle _{A}\left| 1\right\rangle _{B} &\rightarrow&
\Ket{01} \equiv \frac{1}{\sqrt{2}}\left( |01\rangle ^{+}+|01\rangle ^{-}\right) \\
\left| 1\right\rangle _{A}\left| 0\right\rangle _{B} &\rightarrow&
\Ket{10} \equiv \frac{1}{\sqrt{2}}\left( |01\rangle ^{+}-|01\rangle ^{-}\right) \\
\left| 1\right\rangle _{A}\left| 1\right\rangle _{B} &\rightarrow&
|11\rangle ^{+}.
\end{eqnarray}
\end{subequations}
Note that the two particle states at the r.h.s of (\ref{9}) have a trivial 
evolution at the trapping frequency (or multiples of it) 
that can be removed by including this phase
in the definition of the single particle states. 

We will take states (10) as the starting set for the gate operation and, 
after setting the initial state, we will adiabatically realize the gate. 
In this adiabatic regime, if we start in an energy eigenstate 
the system will follow this time-dependent energy eigenstate during the whole gate process. 
The only allowed transitions are those corresponding to
states that (i) are initially degenerate in energy, and, at short distances, 
(ii) become coupled via tunneling and/or cold collisions. Therefore,
in order to find the most suitable gate to be implemented in this system, 
we have to identify these resonant couplings.    

For this aim we will first discuss the ideal case for which there is no interaction
between the atoms, i.e.\ the case where $a_t =0$ in (\ref{scat}). We then have the
following resonant couplings:
\begin{subequations}
\label{10}
\begin{eqnarray}
\Ket{00}^+ &\leftrightarrow &|\widetilde{00}\rangle ^{+} \\
\Ket{01}  & \leftrightarrow &  \Ket{\widetilde{01}}  \leftrightarrow  \Ket{10}
\leftrightarrow  \Ket{\widetilde{10}}  \leftrightarrow  \Ket{01}\\
\Ket{11}^+ &\leftrightarrow &|%
\widetilde{11}\rangle ^{+}
\end{eqnarray}
where $|\widetilde{01}\rangle \equiv \frac{1}{\sqrt{2}}(|\widetilde{01}%
\rangle ^{+}+|\widetilde{01}\rangle ^{-})$ and $|\widetilde{10}\rangle
\equiv \frac{1}{\sqrt{2}}(|\widetilde{01}\rangle ^{+}-|\widetilde{01}\rangle
^{-})$. Therefore, there is a non-negligible probability (even if we move
the two traps adiabatically) to have both atoms in the
same trap after the gate operation. Note that the kinetic and trapping terms
of the Hamiltonian do not directly couple $\left| 01\right\rangle $ with 
$\left| 10\right\rangle$ since they are single particle Hamiltonians and, 
therefore, they do not allow for the simultaneously change of the motional 
states of both atoms. 
The coupling between $\left| 01\right\rangle $ and 
$\left| 10\right\rangle $ is mediated through the double occupancy states 
$|\widetilde{01}\rangle $ and $| \widetilde{10}\rangle $. 
Clearly, in the non-interacting case, a quantum gate operation always has to face
with double-occuppancy which makes the problem hard to handle. 

Figures 2(a) and (b) show, for a particular parameter set, 
the final state of the system after the whole
process of approaching and separating the traps as a function of the
scattering length. In (a) the initial state is $\left| 01\right\rangle$ and in (b) 
$\left| 11\right\rangle^+$. 
Although the scattering length has a constant value that depends on the 
atom under consideration, 
it is used in this plot as a free parameter to illustrate 
the double-occupancy problem. Notice that by changing $\omega_p$ it is possible to
tune the strength of the effective interaction potential, cf.\ Eq.\ (\ref{scat}).
Figs.~2(a) and (b) clearly show that, for $a_t=0$, double-occupancy is
indeed very important in the final state of the system.

\begin{figure}
%
%
\begin{center} 
\includegraphics[width=8.8cm]{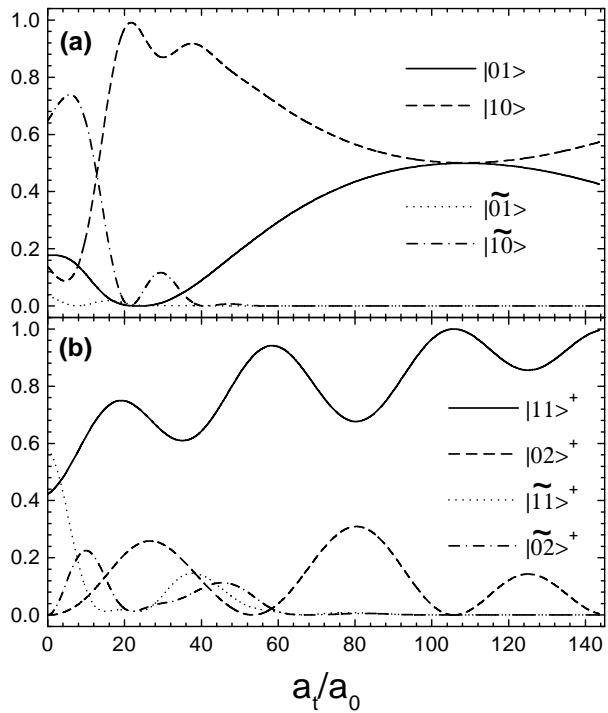} 
\end{center}
\label{scattering}
\caption{
Populations of the final state of the system after adiabatically approaching
 and separating the traps as a function
of the scattering length. The initial state is 
(a) $\Ket{01}$ and (b) $\Ket{11}^+$, respectively. 
The parameter setting is: $\omega_x= 1.25 \times 10^4$ s$^{-1}$,
$\omega_p= 7.9 \times 10^6$ s$^{-1}$, $1/\alpha = 241$~nm,
$a_{\max }\alpha =5 $, $a_{\min }\alpha =1.99 $, $\omega_x t_{r}=70$, 
and $\omega_x t_{i}=69$. 
}
\end{figure}

The problem of double-occupancy is naturally suppressed when one takes into
account the interaction between the atoms. In this case, double-occupancy
states are no longer degenerate with single-occupancy states and we can
neglect the probability to find double occupation in the final state
by adiabatically moving the traps.
Thus, in the presence of interaction, the resonant couplings read
\end{subequations}
\begin{subequations}
\label{11}
\begin{eqnarray}
\left| 01\right\rangle  &\leftrightarrow
&\left| 10\right\rangle  \\
\left| 11\right\rangle^+ &\leftrightarrow
&|02\rangle ^{+}\label{unwanted}
\end{eqnarray}
where
$|02\rangle ^{+}=
\left( \left| \overline{0}\right\rangle
_{L}\left| \overline{2}\right\rangle _{R}+\left| \overline{2}\right\rangle
_{R}\left| \overline{0}\right\rangle _{L}+
\left| \overline{0}\right\rangle _{R}\left| \overline{2}
\right\rangle _{L}+\right.$ $ \left. \left| \overline{2}\right\rangle _{L}\left| \overline{0}%
\right\rangle _{R}\right)/2.$ 
Notice that now the collisional interaction term (\ref{scat}) allows for the simultaneous
change of the motional states of both atoms. 
The role of these couplings is clearly shown in Fig.~2.
In Fig.~2(a), where the initial state is $\Ket{01}$, 
double-occupancy populations in the final state start to decrease and eventually
vanish as soon as the scattering length is increased. 
When the initial state is $\Ket{11}^+$, Fig.~2(b), double-occupancy 
also vanishes as the scattering length increases,
but then the population of state $\Ket{02}^+$ becomes important.

Therefore, the coupling given in Eq.~(12a) suggests the implementation 
of a \sqrtsw gate, as long
as we are able to suppress/control coupling (\ref{unwanted}). The degeneracy
between $\left| 1 1\right\rangle ^{+}$ and $|02\rangle ^{+}$ 
can be broken, for instance, by taking an anharmonic trapping potential
such that the vibrational frequencies are no longer equally spaced. 
In addition, it is possible to
adjust the interaction time in such a way that, at the end of the gate
operation, state $|02\rangle ^{+}$ is not populated. In what follows, 
we will focus on this last possibility.

\section{\sqrtsw gate}

The \sqrtsw gate has the following effect on the states of the computational
basis:
\end{subequations}
\begin{subequations}
\label{table}
\begin{eqnarray}
\left| 0\right\rangle _{A}\left| 0\right\rangle _{B} &\rightarrow &\left|
0\right\rangle _{A}\left| 0\right\rangle _{B}  \\
\left| 0\right\rangle _{A}\left| 1\right\rangle _{B} &\rightarrow &
\frac{ 1+i}{2} \left| 0\right\rangle _{A}\left| 1\right\rangle
_{B}+\frac{1-i}{2} \left| 1\right\rangle _{A}\left| 0\right\rangle _{B}%
 \label{01undersqrtsw}\\
\left| 1\right\rangle _{A}\left| 0\right\rangle _{B} &\rightarrow &
\frac{1-i}{2} \left| 0\right\rangle _{A}\left| 1\right\rangle
_{B}+\frac{1+i}{2} \left| 1\right\rangle _{A}\left| 0\right\rangle _{B} \\
\left| 1\right\rangle _{A}\left| 1\right\rangle _{B} &\rightarrow &\left|
1\right\rangle _{A}\left| 1\right\rangle _{B}  
\end{eqnarray}
\end{subequations}
It is straightforward to check that the successive 
application of two \sqrtsw gates exchanges the states of the qubits, i.e., 
$\uswap= \usqrtsw \cdot \usqrtsw $. As it has been mentioned before, 
the \sqrtsw gate together with single qubit operations suffices
to realize any quantum algorithm \cite{swap} which is not the
case for the SWAP gate itself. 
A simple way to prove this, consists of showing 
that the universal CNOT gate can be obtained from \sqrtsw gates 
and single-qubit operations. In fact, a possible sequence is (see Appendix B): 
\begin{equation}
U_{\text{CNOT}}=  
{H}_{A} \sigma_{A}^{-1} \sigma_{B} 
\usqrtsw  \sigma_{A}^{2}
\usqrtsw {H}_{A} 
\end{equation}
where $ {H}_{A}$ and $ {\sigma }_{A,B}$ are 
single-qubit operations.
Additionally, sequences involving single-qubit operation exclusively 
on one of the qubits, e.g., only on A, can be realized \cite{swap}.

\subsection{Gate simulation}

To simulate the gate operation, we have numerically integrated 
the time-dependent Schr\"odinger equation for the Hamiltonian given in (3) 
and (5) with the two-particle wave-function expanded in the previously 
introduced two-particle basis.  
Figures 3(a)-(c) show the result of a \sqrtsw gate operation for a 
scattering length of $a_t=106$~$a_0$ corresponding to $^{87}$Rb
atoms in the spin triplet. The parameter setting is as in Fig.~2
and the initial state is (a) $\Ket{00}^+$, (b) 
$\Ket{01}$, and (c) $\Ket{11}^+$. 
The parameter values are chosen to reproduce the gate operation
given in (\ref{table}) as well as to suppress the $\Ket{02}^+$ 
population in the outgoing state of Fig.~3(c). 
Notice that states representing double occupation are populated at
close distances for all three cases. However, these populations vanish
after the eventual separation of the traps since the traps
are moved adiabatically and single-occupancy states are not 
degenerate with double-occupancy ones.

\begin{figure}
%
\begin{center}
\includegraphics[width=8.8cm]{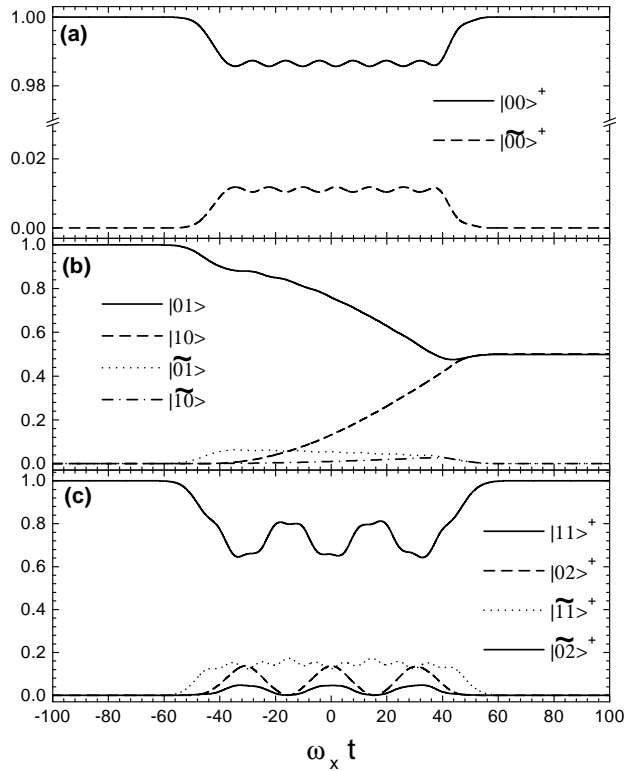}  
\end{center}
\label{gate}
\caption{
Simulated \sqrtsw gate operation for $^{87}$Rb with $a_t=106$~$a_0$.
The rest of parameters as in Fig.~2.
The initial state of the system is (a) $\Ket{00}^+$, (b) $\Ket{01}$,
and (c) $\Ket{11}^+$. We note that for (b) the final relative phases of
$\Ket{01}$ and $\Ket{10}$ are as in Eq.~(\ref{01undersqrtsw}).
}
\end{figure}

For $^{85}$Rb with negative scattering length it is slightly more involved
to find parameters for the gate realization since, due to the attractive character
of the interaction, double occupation states can more easily become resonant to
single occupation states, e.g., $\Ket{\widetilde{01}}^+$ with $\Ket{00}^+$. 
The parameters must be chosen to avoid this degeneracy between double and single occupation
states. Fig.~4 shows the result of a gate simulation for $^{85}$Rb. Unlike for 
$^{87}$Rb, Fig.~4(a) shows that starting from $\Ket{00}^+$ state
$\Ket{\widetilde{01}}^+$ is populated during the gate operation.

On the other hand, it is important to notice that the results obtained for $^{87}$Rb
(Fig. 3)
can be also directly implemented in $^{85}$Rb by making use of the strong variation of the 
scattering length in the vicinity of a magnetic field induced Feschbach resonance 
\cite{Feschbach}.

\begin{figure}
%
\begin{center}
\includegraphics[width=8.8cm]{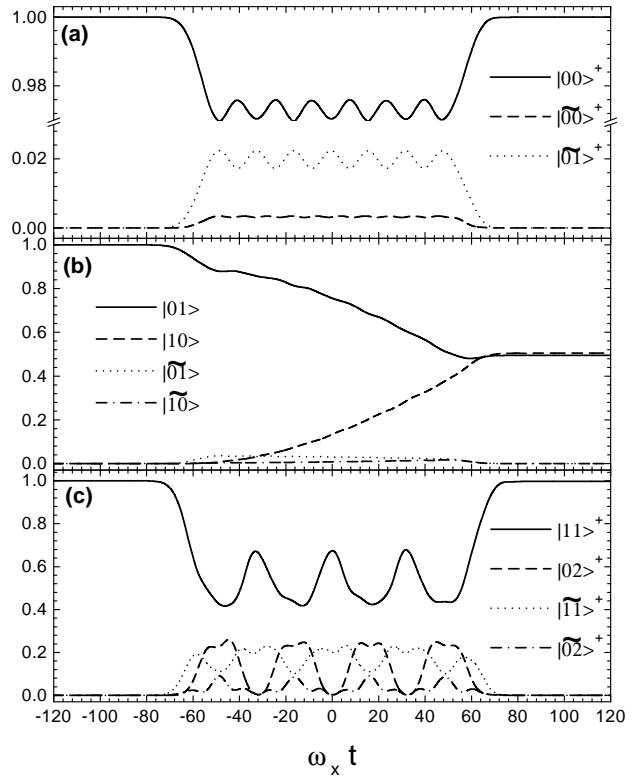}  
\end{center}
\label{gate2}
\caption{
Simulated \sqrtsw gate operation for $^{85}$Rb and the following parameter values:
$\omega_x= 1.25 \times 10^4$ s$^{-1}$, 
$\omega_p= 1.6 \times 10^6$ s$^{-1}$, $1/\alpha = 244$~nm,
$a_{\max }\alpha =5 $, 
$a_{\min }\alpha =1.956 $, $\omega_x t_{r}=77$, 
$\omega_x t_{i}=97.2$, and $a_t=-369$~$a_0$.
The initial state of the system is (a) $\Ket{00}^+$, (b) $\Ket{01}$,
and (c) $\Ket{11}^+$.
}
\end{figure}

To check the accuracy of the previous simulations in which the two-particle
wavefunction was expanded in a finite set of states, we also have
numerically integrated the Schr\"odinger equation for the two-particle
spatial wavefunction by using an operator split method and an FFT routine. 
Fig.~5 shows the results of this integration for the same 
parameter values as in Fig.~3 \cite{ourweb}. The snapshots give the
joint-probability distributions for the two particles 
for three different initial states: (a) $\Ket{00}^+$, 
(b) $\Ket{01}$ and (c) $\Ket{11}^+$. 
The bosonic nature of the atoms manifests in the symmetry of the joint-probability
distribution along the diagonal $x_1=x_2$.
In (a) and (c) the final state
coincides with the initial one in accordance with Eqs.\ (\ref{table}). 
In (b) $\Ket{01}$ evolves towards the maximally entangled state 
$\left[(1+i)\Ket{01}+(1-i)\Ket{10}\right] /2$ 
whose joint-probability distribution corresponds to the donut-like
shape of the last frame.

\begin{figure}
%
%
\begin{center}
\includegraphics{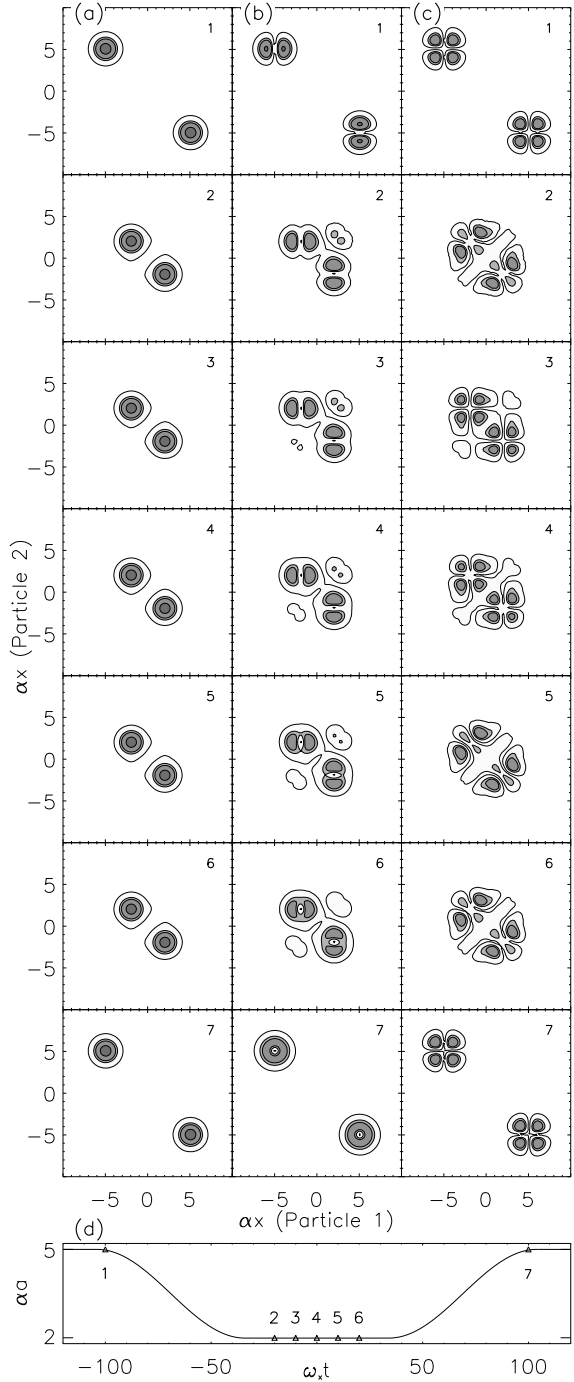} 
\end{center}
\label{jointprop}
\caption{Snapshots of the spatial two-particle wavefunction
$|\psi(x_1,x_2)|^2$ for $^{87}$Rb.
The parameters
are as in Fig.\ 3. The horizontal and vertical axes of
each plot show the coordinate of the first and second particle, respectively.
Initially there is one particle in each of the traps. 
The symmetry along the diagonal $x_1=x_2$ is due to the bosonic statistics.
A particle in the ground state of one of the traps corresponds to a
gaussian distribution in the direction of the respective axis while
one node corresponds to the first excited state.
Thus the initial states are (a) $\Ket{00}^+$, (b) $\Ket{01}$,
and (c) $\Ket{11}^+$. The time for the snapshots is shown in (d).
See \protect\cite{ourweb} for animated illustrations of the gate operation.
}
\end{figure}

The accuracy of the simulated gate operation $U$
with respect to the perfect gate operation $U_{\text{\sqrtsw}}$
as given by Eqs.~(\ref{table}) is computed through the averaged fidelity, i.e.\  
\begin{equation}
F=\overline{\text{Tr} [{U}\rho {U}^{\dagger}\,
U_{\text{\sqrtsw}}\rho U_{\text{\sqrtsw}}^{\dagger}]}
\end{equation}
where the average is taken over the four orthogonal pure input states
 $\rho$ from Eqs.~(\ref{table}).
Figure 6(a) shows for $^{87}$Rb the averaged fidelity $F$ of the gate
in the parameter plane $t_r$ versus $a_{\min}$.
The rest of the parameters are as in Fig.~3. Clearly, the fidelity is very
sensitive to the minimum distance due to the exponential dependence
of tunneling at this distance. Note that the fidelity of the
gate operation corresponding to the parameters of Fig.~3 is $F>0.9997$
with a gate duration of $2t_r + t_i \sim 17$~ms for $\omega_x=1.25 \times 10^4$~s$^{-1}$.

\begin{figure}
%
%
\begin{center}
\includegraphics[width=8.5cm]{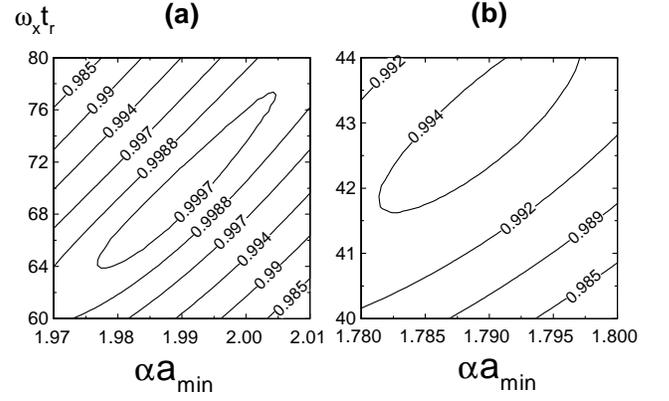} 
\end{center}
\label{fidelity}
\caption{Averaged fidelity of the gate operation in the parameter plane $t_r$ versus $a_{\min}$.
The interaction time is (a) $\omega_x t_i=69$ and (b) $\omega_x t_i=20$. 
The rest of the parameters as in Fig.~3.
}
\end{figure}

An important issue is how much the gate duration can be decreased
while maintaining a high fidelity. In Fig.~6(b) the gate duration is reduced 
by a factor of two which reduces the fidelity by around one order of magnitude. 
In fact, as soon as the rising time $t_r$ is decreased, 
non-adiabatic effects occur which in turn result in the population of several 
unwanted states, e.g.\ double-occupation, in the final state of the system.  
However, it could be possible to use the techniques developed in \cite{optimization}
to optimize the speed of the gate operation, while suppressing 
excitations to these unwanted states. 

\subsection{Quantum correlations}

Let us briefly consider entanglement in the context of the
\sqrtsw gate. As already discussed the initial and, as long as double
occupancy is suppressed, also the final state consist of well separated
and thus
for practical purposes distinguishable particles, such that the usual
notions of entanglement apply. Let us consider in particular the
separable state $\Ket{01}$ that under the \sqrtsw operation
evolves to the maximally entangled state from Eq.\ (\ref{01undersqrtsw}).
During the operation bosonic statistics become important
and thus we have to distinguish between statistical correlations
arising from symmetrization on one hand and quantum correlations
which are useful in the context of quantum information on the other hand. Such a
distinction has been discussed in \cite{dots2,john+maciej} for 
fermionic two-particle states where the notion of Slater rank  was
introduced and a fermionic measure of entanglement was derived. In 
\cite{dots2} these methods have also been used to study correlations
in the context of a quantum gate operation for two electrons in 
quantum dots. They have been translated to bosons
in \cite{john+kai} and moreover a bosonic von Neumann
entropy has been defined in \cite{paskauskas}. Here we want to study
to which extent these techniques can be applied to our particular
gate operation.

Let us 
write a general two-boson state in an $N$-dimensional single particle 
space as $\Ket{v}=\sum_{i,j=1}^N v_{ij}\bc{i}\bc{j}\Ket{\0}$
where $\bc{i}$ and $b_i$ are bosonic creation and annihilation operators
and $\Ket{\0}$ is the vacuum state
such that $\bc{i}\Ket{\0}=\Ket{\bar i}$. The complex symmetric matrix
$v_{ij}=v_{ji}$ is normalized as $\tr(v^{\dagger}v)=1/2$. If new bosonic
annihilation operators $\bas{i}=\sum_{ij}U_{ij}\ba{j}$ are introduced by a unitary
transformation $U$ of the single particle space,
then $v$ transforms as $UvU^{T}$. Now we find that for 
every symmetric complex matrix $v$ there exists a unitary $U$ such that
$UvU^{T}$ is diagonal,
i.e.\ $UvU^{T}=\text{diag}[\lambda_1,\ldots,\lambda_r,0,\ldots,0]$
with $\lambda_i>0$ \cite{john+kai,paskauskas}.
Here $r$ is called the {\it Slater rank} of $\Ket{v}$ and
$\Ket{v}=\sum_{i}^r\lambda_i\bcs{i}\bcs{i}\Ket{\0}$ its
{\it Slater decomposition}.

In our case the initial state $\Ket{01}$ has Slater rank two while the
final state
$\usqrtsw \Ket{01}$ has Slater rank four. A bosonic measure of entanglement
has been defined in \cite{john+kai} only for the case of a two-dimensional
single particles space, but in \cite{paskauskas} a bosonic version of the
von Neumann entropy has been proposed that, as a function of the Slater
coefficients $\lambda_i$ reads
\begin{equation}
S_B=-\sum_{i=1}^r \lambda_k^2\log_2(\lambda_k^2).
\end{equation}
$S_B$ ranges from $S_B=0$ for states with Slater rank $1$ to 
$S_B=\log_2(N)$ for states with Slater rank $N$ and all $\lambda_i$
equals. For the case considered here we have $S_B(\Ket{01})=1$ and $S_B(\usqrtsw\Ket{01})=2$.
The function $S_B(t)/2$ is plotted in Fig.~6 
for the complete \sqrtsw
operation (the factor is such that $S_B$ for $N=4$ ranges between $0$ and $1$).
 The figure also shows the product $S\cdot p_{\text{single}}$ where
$S$ is the von-Neumann entropy calculated
by projecting the state onto the space spanned by the set
$\left\{\Ket{00}^+,\,\Ket{01}^+,\,\Ket{01}^-,\,\Ket{11}^+\right\}$
and renormalizing. If now states $\Ket{\bar{i}}_L$ are considered as
being distinguishable from states
$\Ket{\bar{i}}_R$ then $S$ can be calculated as for distinguishable
particles. $p_{\text{single}}$ is the probability to find the state
in the space spanned by the given set.

\begin{figure}
%
%
\begin{center}
\includegraphics[width=8.5cm]{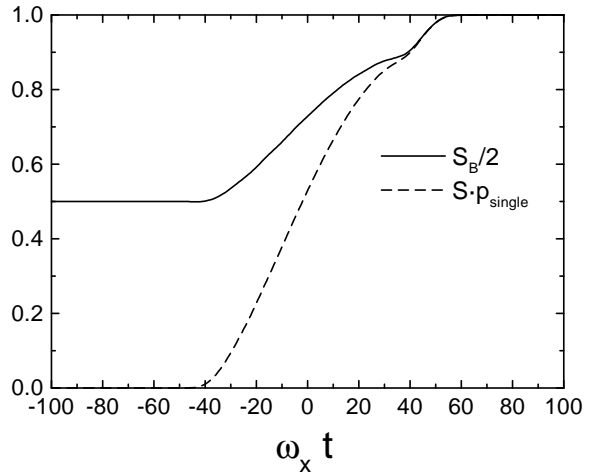} 
\end{center}
\label{entanglement}
\caption{Correlations for the \sqrtsw gate operation applied to $\Ket{01}$.
Bosonic von Neumann entropy $S_B/2$ (solid line) and
von Neumann entropy $S$ (dashed line) after projection onto the space 
spanned by $\left\{\Ket{00}^+,\,\Ket{01}^+,\,\Ket{01}^-,\,\Ket{11}^+\right\}$
multiplied by the probability $p_{\text{single}}$ to find the state
in the space spanned by this set.}
\end{figure}

Fig.~7 shows that in the limit of large
spatial separation of the particles 
neither the Slater rank nor the bosonic von Neumann entropy from
\cite{paskauskas} coincide with the notion of Schmidt rank and
von Neumann entropy for distinguishable particles. There are two reasons
for this obvious difference: (i) Slater rank and bosonic von Neumann
entropy are invariants under arbitrary transformations of the
single particle space, while in our case the partition of the full
single-particle space into the space spanned by $\Ket{\bar{i}}_L$ on
the one hand and $\Ket{\bar{i}}_R$ on the other is rather fixed; (ii)
they do not take into account that here we want to
avoid double-occupation in the final state.

 In \cite{zanardi}
P. Zanardi follows a different approach to calculate an entropy for
indistinguishable particles which consists of mapping the bosonic
(or fermionic) state to an occupation number basis
and calculating the von Neumann entropy in this basis.
If this is applied here then the entropy raises from
$0$ for $\Ket{01}$ to $1$ for $\usqrtsw\Ket{01}$ but in between it 
has a maximum of $\sim1.4$. As can be seen from the analysis in \cite{fisher}
this is due to the presence of spatial entanglement during the operation.
It is however not clear whether this type of entanglement is useful in this
context where the two qubits are explicitely implemented in the vibrational
states of each trap. These results thus call for further
investigation of the problem of quantum correlations in bosonic states.

\section{Practical considerations}

Scalable systems of optical microtraps based on the dipole
force can be realized by focusing a single red-detuned laser beam with
a microlens array \cite{opt1,opt3}.
The temporal evolution of the trap separation as shown in Fig.~1 can be realized
(i) by using two parallel laser beams focused in such a way that the trapping potentials
are longitudinally shifted along the common laser beam direction or
(ii) by illuminating the microlens array with two laser beams under slightly different
angles with the possible inclusion of an additional two-dimensional confining potential
perpendicular to the direction of the trap displacement \cite{opt1}. 
For the parameters we used in the gate simulations, the minimum distance of $\alpha a_{min}\sim 2$ 
corresponds to a separation of the traps of $1$~$\mu$m 
which is achievable in the present optical microtraps \cite{opt3}.
With laser powers of $1$-$10$~${\rm mW}$ per trap,
rubidium atoms can be trapped with typical trapping
frequencies along the laser beam direction of $\omega _{x}\sim 10^{4}$-$%
10^{5}$~${\rm s}^{-1}$ while the transverse trapping frequencies can be 
one or two orders of magnitude larger \cite{opt1}. 
Additionally, sideband cooling
could be applied to cool the atoms to the ground state of each trap in all
dimensions. 
 
In the optical microtrap experiments, 
the trapping potential is gaussian-shaped with typical depths
of 1-10 mK~$\times $~$k_B$ \cite{opt1}. For a single trap it is thus a 
good approximation to assume a harmonic
potential for the lower lying states. For two traps being
close together the actual potentials will deviate from the
form assumed in (\ref{trapx}). Nevertheless, it is possible to generalize 
the methods applied here to these particular potentials. 
To allow for single site addressing, 
typical initial distances between the traps have to be about 5-20~$\mu$m
which for rubidium means $\alpha a_{max} = 10-40$ instead of $\alpha a_{max} = 5$
which we used in our previous calculations. 
In this case, it is straightforward to estimate
that the process of adiabatically approaching the traps
to the interaction distance and eventually separating them again
requires another $\sim10$ ms.
Therefore, the whole gate process could be realized 
in a few tens of milliseconds which is enough for a proof-of-principle experiment. 
In fact, the lifetime of the atoms in the traps is about $100-1000$ ms
while spontaneous scattering occurs in $\sim 10$ ms. 
However for strongly confining trapping potentials only a small
fraction of the occurring spontaneous scattering processes leads
to a change in the vibrational state \cite{echo}.

In a recent paper, E.~Charron {\it et al.}~\cite{williams} 
proposed the realization of a phase gate in an optical lattice
where the qubits were also implemented in the motional states. Two 
linear counterpropagating beams from the fundamental and first harmonic
of a CO$_2$ laser were used to produce an intensity gradient optical lattice.
The barrier between two neighboring traps could be raised or
lowered by changing the intensity ratio between the two beams. We notice
that the realization of a \sqrtsw gate as discussed here should also
be possible in this setup although the implementation in optical
microtraps presents some advantages such as being not sensitive to
the phase fluctuations of the lasers.  
 
\section{Conclusion}

We have investigated quantum computation in optical microtraps 
with the qubits implemented in the motional states of neutral atoms,
and tunneling and cold controlled collisions accounting for the interaction between
two different qubits.
A time-dependent two-particle orthogonal basis has been introduced
to simulate the gate operation and to compute entanglement 
throughout the whole gate process. The bosonic statistic nature of the particles
and its role in entanglement has been discussed in detail. 
We have demonstrated the capability 
of optical microtraps to realize a high-fidelity \sqrtsw gate operation
in the few tens of milliseconds range. Finally, some practical considerations
for the physical implementation of this quantum gate have been discussed.

\section*{Acknowledgments}

This work was supported by the European Commission through the projects 
EQUIP and ACQUIRE within the framework of the
IST-program, as well as through a Marie Curie Fellowship (J.M.) 
and by the Deutsche Forschungsgemeinschaft through the research
program 'Quanteninformationsverarbeitung' and SFB 407.
We thank A.~Sanpera, W.~Ertmer, W.~Lange, C.~Williams, R.~Dumke, F.~Hulpke, P.~Hyllus,
O.~G\"{u}hne, J.~Korbicz, T.~M\"{u}ther, L.~Santos, T.~Schulte, and M.~Volk
for helpful discussions.  
The authors are solely responsible for
information communicated and the European Commission is not responsible for
any view or results expressed.

\begin{appendix}
\section{Gram-Schmidt orthonormalization}\label{gmmethod}

In this appendix we will show how to construct the time-dependent
orthonormal single-particle states, denoted by
$\left| \overline{i}\right\rangle 
 _{L}$ and $\left| \overline{i}\right\rangle _{R}$, from the harmonic
oscillator energy eigenstates $\left| i\right\rangle _{L}$ and $\left|
i\right\rangle _{R}$ for the left and the right trap, respectively.
We start by defining states involving one state of each trap: 
\begin{equation}
\left| i\right\rangle ^{\pm } \equiv 
\frac{1}{\sqrt{2}}\left[ \left| i\right\rangle
_{L}\pm (-1)^{i}\left| i\right\rangle _{R}\right] \quad \quad
i=0,1,2,3,\ldots
\end{equation}
where the superscript $+$ ($-$) indicates positive (negative) parity with
respect to the middle between the two traps. We then group these
states according to their parity in two sets 
$S^{\pm }=\left\{ \left| 0\right\rangle
^{\pm },\left| 1\right\rangle ^{\pm },...\right\} $ and focus first on the
positive parity set $S^{+}$. This set contains states that are neither
orthogonal nor normalized. To perform the orthonormalization, we
use the Gram-Schmidt (GM) method starting with the following normalized
function: 
\begin{equation}
\phi _{0}^{+}(x,a)\equiv \frac{\left\langle x\right. \left| 0\right\rangle
^{+}}{\int \big| \left\langle x\right. \left| 0\right\rangle ^{+}\big|
^{2}dx},
\end{equation}
Then, we define the first linearly independent function $\phi _{1}^{+}$ as:
\begin{equation}
\phi _{1}^{+}(x,a)=\frac{\left\langle x\right. \left| 1\right\rangle
^{+}+a_{10}\phi _{0}^{+}(x,a)}{\int \big| \left\langle x\right. \left|
1\right\rangle ^{+}+a_{10}\phi _{0}^{+}(x,a)\big| ^{2}dx},
\end{equation}
where $a_{10}=-\int \phi _{0}^{+}(x,a)\left\langle x\right. \left| 1\right\rangle
^{+}dx$ which guaranties $\left\langle \phi _{0}^{+}\right. \left| \phi
_{1}^{+}\right\rangle =0$. We repeat this procedure to obtain the rest
of the linearly independent functions $\phi _{2}^{+},\phi
_{3}^{+},...$ with positive parity. In an analogous way, we
determine from $S^{-}$ the set of linearly independent
functions $\left\{ \phi _{0}^{-},\phi _{1}^{-},\phi _{2}^{-},\phi
_{3}^{-},...\right\}$. An important feature of the GM method when
applied to a set of states with the same parity is that the
constructed orthonormal states retain the parity of the original
set of states. Thus states from $\left\{\phi_i^+\right\}$ and
$\left\{\phi_i^+\right\}$ have positive and negative parity, respectively
and, therefore, the whole set  $\left\{ \phi _{0}^{\pm },\phi _{1}^{\pm
},\phi _{2}^{\pm },\phi _{3}^{\pm },...\right\} $ is orthonormal.
Explicitly, the first four orthonormalized functions read 
\begin{subequations}
\begin{eqnarray}
\phi _{0}^{\pm }(x,a) &=&\xi _{0}^{\pm }(a)\left\langle x\right. \left|
0\right\rangle ^{\pm }, \\
\phi _{1}^{\pm }(x,a) &=&\xi _{1}^{\pm }(a)\left( \left\langle x\right.
\left| 1\right\rangle ^{\pm }\pm \frac{x\alpha ^{3/2}}{\sqrt[4]{4\pi }}%
e^{-a^{2}\alpha ^{2}}\left\langle x\right. \left| 0\right\rangle ^{\mp
}\right) ,
\end{eqnarray}
\end{subequations}
where 
\begin{subequations}
\label{A5}
\begin{eqnarray}
\xi _{0}^{\pm }(a) &=&\frac{1}{\sqrt{1\pm e^{-a^{2}\alpha ^{2}}}},  \\
\xi _{1}^{\pm }(a) &=&\frac{e^{a^{2}\alpha ^{2}}}{\sqrt{\left(
e^{a^{2}\alpha ^{2}}\pm 1\right) \left( e^{a^{2}\alpha ^{2}}-e^{-a^{2}\alpha
^{2}}\pm 2a^{2}\alpha ^{2}\right) }}.
\end{eqnarray}
For the sake of brevity, we do not explicitly show the analytical
expressions for the rest of the $\phi_i^{\pm}$.

Once we have obtained the orthonormal set $\{\phi _{i}^{\pm }\}$%
, it is straightforward to write down the single-particle basis that
we will use: 
\end{subequations}
\begin{subequations}
\label{spbasisstates}
\begin{eqnarray}
\left\langle x\right. \left| \overline{i}\right\rangle _{L} &=&\frac{1}{%
\sqrt{2}}\left( \phi _{i}^{+}+\phi _{i}^{-}\right) \\
\left\langle x\right. \left| \overline{i}\right\rangle _{R} &=&\left(
-1\right) ^{i}\frac{1}{\sqrt{2}}\left( \phi _{i}^{+}-\phi _{i}^{-}\right) , 
\end{eqnarray}
\end{subequations}
These states are orthonormal due to the orthonormality of the $\phi _{i}^{\pm }$
and in the limit $a\alpha \gg 1$ become the corresponding harmonic oscillator energy eigenstates for
each trap. These new orthonormal states do not have in general a well
defined parity with respect to the center of the corresponding trap but it is
straightforward to check from Eqs.\ (\ref{spbasisstates}) that they satisfy the following property
under parity transformation with respect to the middle of the traps:
\begin{equation}
\langle x|\overline{i}\rangle _{L,R}\mapsto \left( -1\right) ^{i}\langle x|%
\overline{i}\rangle _{R,L}.
\end{equation}

\section{Universality of the \sqrtsw gate}

Our goal here is to write down the sequence of steps required to build the
CNOT gate, which, in the computational basis
$\Ket{0}_A\Ket{0}_B$, $\Ket{0}_A\Ket{1}_B$, $\Ket{1}_A\Ket{0}_B$ and
$\Ket{1}_A\Ket{1}_B$, reads 
\begin{equation}
\ucnot=\left( 
\begin{array}{cccc}
1 & 0 & 0 & 0 \\ 
0 & 1 & 0 & 0 \\ 
0 & 0 & 0 & 1 \\ 
0 & 0 & 1 & 0
\end{array}
\right) ,\label{B1}
\end{equation}
from the \sqrtsw gate  
\begin{equation}
\usqrtsw=\left( 
\begin{array}{cccc}
1 & 0 & 0 & 0 \\ 
0 & \frac{1+i}{2} & \frac{1-i}{2} & 0 \\ 
0 & \frac{1-i}{2} & \frac{1+i}{2} & 0 \\ 
0 & 0 & 0 & 1
\end{array}
\right) .
\end{equation}
The single qubit operations we need are on one hand the Hadamard gate 
\begin{equation}
 {H}=\frac{1}{\sqrt{2}}\left( 
\begin{array}{cc}
1 & 1 \\ 
1 & -1
\end{array}
\right) ,
\end{equation}
and on the other hand the following combination of identity and Pauli
$\sigma_z$ matrices:

\begin{equation}
 {\sigma }=\left( 
\begin{array}{cc}
1 & 0 \\ 
0 & -i
\end{array}
\right) =e^{-i\frac{\pi }{4} {I}}e^{-i\frac{\pi }{2} {\sigma }%
_{z}}.
\end{equation}
Let us call $ {H}_{A,B}$ and $ {\sigma }_{A,B}$ the
corresponding single-qubit operations for qubit A or B. 
Now it is easy to check that the
following combination of single-qubit operations and \sqrtsw gates 
yields the phase gate: 
\begin{eqnarray}
\uphase&=&
{\sigma}_{A}^{-1} \sigma_B
\usqrtsw  {\sigma }_{A}^{2}\usqrtsw  \label{phase}\\
&=& \left( 
\begin{array}{cccc}
1 & 0 & 0 & 0 \\ 
0 & 1 & 0 & 0 \\ 
0 & 0 & 1 & 0 \\ 
0 & 0 & 0 & -1 
\end{array}
\right)\nonumber
\end{eqnarray}
This sequence is not unique, and more sophisticated sequences involving
single-qubit operations on only one of the qubits can be implemented \cite
{swap}. Finally, to obtain the CNOT gate it is enough to apply a Hadamard
gate on the qubit A at both sides of Eq.\ (\ref{phase}), i.e.
\begin{equation}
\ucnot= {H}_{A}\uphase {H}_{A}.
\end{equation}

\end{appendix}


\begin{thebibliography}{3}


\bibitem{NMR} D. G. Cory, A. F. Fahmy, T. F. Havel, Proc. Natl. Acad. Sci.
USA {\bf 94}, 1634 (1997); N. A. Gershenfeld, I. L. Chuang, Science {\bf 275}, 350
(1997); I. L. Chuang, N. Gershenfeld, M. G. Kubinec, and D. W. Leung, Proc.
R. Soc. London A {\bf 454}, 447 (1998).

\bibitem{ion}  J.I. Cirac and P. Zoller, Phys. Rev. Lett. {\bf 74}, 4091
(1995); J. I. Cirac and P. Zoller, Nature (London) {\bf 404}, 579 (2000);
T. Calarco, J. I. Cirac, and P. Zoller, Phys. Rev. A {\bf 63}, 062304 (2001).

\bibitem{QED}  A. Barenco, D. Deutsch, A. Ekert, and R. Jozsa, Phys. Rev.
Lett. {\bf 74}, 4083 (1995); T. Sleator and H. Weinfurter, Phys. Rev. Lett. 
{\bf 74}, 4087 (1995); Q. A. Turchette, C. J. Hood, W. Lange, H. Mabuchi, and H.
J. Kimble. Phys. Rev. Lett. {\bf 75}, 4710 (1995); P. Domokos, J. M. Raimond,
M. Brune, and S. Haroche, Phys. Rev. A {\bf 52}, 3554 (1995);
T. Pellizzari, S. A. Gardiner, J. I. Cirac and  P. Zoller, Phys. Rev. Lett. {\bf 75},
3788 (1995);
L. G. Lutterbach and L. Davidovich, Opt. Expr. {\bf 3}, 147 (1998);
V. Giovanetti, D.Vitali, P. Tombesi, and A. Ekert, Phys. Rev. A {\bf 62}, 032306 (2000). 

\bibitem{swap}  D. Loss and D. P. DiVincenzo, Phys. Rev. A {\bf 57}, 120
(1998).

\bibitem{dots1} G. Burkard, D. Loss, D. P. DiVincenzo, Phys. Rev. B {\bf 59}, 2070
(1999); G. Burkard, G. Seelig, and D. Loss, Phys. Rev. B {\bf 62}, 2581
(1999).

\bibitem{dots2}  J. Schliemann, D. Loss, A. H. MacDonald, Phys. Rev. B {\bf %
63}, 085311 (2001).

\bibitem{na1}  D. Jaksch, H. J. Briegel, J. I. Cirac, C. W. Gardiner, P.
Zoller, Phys. Rev. Lett. {\bf 82}, 1975 (1999).

\bibitem{na2}  T. Calarco, E. A. Hinds, D. Jaksch, J. Schmiedmayer, J. I.
Cirac, and P. Zoller, Phys. Rev. A {\bf 61}, 022304 (2000).

\bibitem{na3}  G. K. Brennen, C. M. Caves, P. S. Jessen, and I. H. Deusch, Phys.
Rev. Lett. {\bf 82}, 1060 (1999).

\bibitem{na4}  D. Jaksch, J. I. Cirac, P. Zoller, S. L. Rolston, R. C\^{o}te
and M. D. Lukin, Phys. Rev. Lett. {\bf 85}, 2208 (2000).

\bibitem{na5} I. E. Protsenko, G. Reymond, N. Schlosser, and P. Grangier, 
Phys. Rev. A {\bf 65}, 052301 (2002).

\bibitem{qedExp} 
E. Hagley, X. Ma{\^\i}tre, G. Nogues, C. Wunderlich, M. Brune, J. M.
Raimond, and S. Haroche, Phys. Rev. Lett. {\bf 79}, 1 (1997);
A. Rauschenbeutel, G. Nogues, S. Osnaghi, P. Bertet, M. Brune, J. M. Raimond,
and S. Haroche, Phys. Rev. Lett. {\bf 83}, 5166 (1999);
J. M. Raimond, M. Brune, and S. Haroche, Rev. Mod. Phys. {\bf 73}, 565 (2001).

\bibitem{ionExp}  C. Monroe, D. M. Meekhof, B. E. King, W. M. Itano, and D. J.
Wineland, Phys. Rev. Lett. {\bf 75}, 4714 (1995); Q. A. Turchette, C.S. Wood,
B.E. King, C. J. Myatt, D. Leibfried, W. M. Itano, C. Monroe, and D. J.
Wineland, Phys. Rev. Lett. {\bf 81}, 3631(1998);

\bibitem{NMRexp} L. M. K. Vandersypen, M. Steffen, G.
Breyta, C. S. Yannoni, R. Cleve, and I. L. Chuang, Phys. Rev. Lett. {\bf 85}%
, 5452 (2000); L. M. K. Vandersypen, M. Steffen, G. Breyta, C. S. Yannoni,
M. H. Sherwood, and I. L. Chuang, Nature {\bf 414}, 883 (2001).

\bibitem{BEC}  They allowed, for instance, to achieve Bose-Einstein condensation
in trapped alkali gases, see 
M. H. Anderson, J. R. Ensher, M. R. Matthews, C. E. Wieman, and
E. A. Cornell, Science {\bf 269}, 198 (1995); C. C. Bradley, C. A. Sackett,
J. J. Tollett, and R.G. Hulet, Phys. Rev. Lett. {\bf 75}, 1687 (1995); K. B.
Davies, M. O. Mewes, M. R. Andrews, N. J. van Druten, D. S. Durfee, D. M.
Durn, and W. Ketterle, Phys. Rev. Lett. {\bf 75}, 3969 (1995).

\bibitem{latt}  See, for instance, I. H. Deutsch and  P. S. Jessen, Phys. Rev. A {\bf 57}, 1972 (1998); and, G.Grynberg and C. Robbilliard,
Phys.Rep. {\bf 355}, 355 (2001) and references therein.

\bibitem{single} D. Frese, B. Ueberholz, S. Kuhr, W. Alt, D. Schrader,
 V. Gomer, and D. Meschede, Phys. Rev. Lett. {\bf 85}, 3777 (2000);
N. Schlosser, G. Reymond, I. Protsenko, and P. Grangier, Nature {\bf 411},
1024 (2001).

\bibitem{mag}  J. Schmiedmayer, Phys. Rev. A {\bf 52}, R13 (1995); J. D.
Weinstein, and K. Libbrecht, Phys. Rev. A {\bf 52}, 4004 (1995). J. Reichel, W.
H\"ansel, and T. W. H\"ansch, Phys. Rev. Lett. {\bf 83}, 3398 (1999); N. H.
Dekker, C. S. Lee, V. Lorent, J. H. Thywissen, S. P. Smith, M. Drndi\'{e},
R. M. Westervelt, and M. Prentiss, Phys. Rev. Lett. {\bf 84}, 1124 (2000);
R. Folman, P. Kruger, D. Cassettari, B. Hessmo, T. Maier, and J.
Schmiedmayer, Phys. Rev. Lett. {\bf 84}, 4749 (2000).

\bibitem{opt1}  G. Birkl, F. B. J. Buchkremer, R. Dumke, and W. Ertmer,
Optics Comm. {\bf 191}, 67 (2001).

\bibitem{opt2}  F. B. J. Buchkremer, R. Dumke, M. Volk, T. M\"uther, G.
Birkl, and W. Ertmer, Laser Physics {\bf 12}, 736 (2002).

\bibitem{opt3}  R. Dumke, M. Volk, T. M\"uther, F. B. J. Buchkremer, G.
Birkl, and W. Ertmer, Pre-print quant-ph/0110140 at xxx.lanl.gov (2001).

\bibitem{divi} D. P. Divicenzo, Fortschr. Phys. {\bf 48}, 771 (2000).

\bibitem{CNOT} J. Gruska, Quantum Computing, London, 1999;
M.A. Nielsen and I.L. Chuang, Quantum Computation and Quantum Information,
Cambridge, 2000.



\bibitem{holonomic}  A. Recati, T. Calarco, P. Zanardi, J. I. Cirac, and P.
Zoller, arXiv:quant-ph/0204030 (2002).

\bibitem{Andersson1}  E. Andersson, M. T. Fontenelle, and S. Stenholm, Phys.
Rev. A {\bf 59}, 3841 (1999).

\bibitem{Andersson2}  E. Andersson and S. Stenholm, Optics Comm. {\bf 188},
141 (2001).

\bibitem{vib}  I. Bouchoule, H. Perrin, A. Kuhn, M. Morinaga, and C.
Salomon, Phys. Rev. A {\bf 59}, R8 (1999); M. Morinaga, I. Bouchoule, J.-C.
Karam, and C. Salomon, Phys. Rev. Lett. {\bf 83}, 4037 (1999).

\bibitem{williams}  E. Charron, E. Tiesinga, F. Mies, and C. Williams, Phys.
Rev. Lett. {\bf 7}, 077901 (2002).

\bibitem{peresbook} A. Peres, Quantum Theory: Concepts and Methods, Dordrecht, 1995.

\bibitem{Feschbach} S. L. Cornish, N. R. Claussen, J. L. Roberts, E. A. Cornell, and C. E. 
Wieman, Phys. Rev. Lett. {\bf 85}, 1795 (2000). 

\bibitem{ourweb} An animated illustration of the \sqrtsw gate operation can be
found at\\
{\em http://www.itp.uni-hannover.de/$\sim$eckert/na/index.html}.

\bibitem{optimization}  W. H\"{a}nsel, J. Reichel, P. Hommelhoff, and T. W.
H\"{a}nsch, Phys. Rev. A {\bf 64}, 063607 (2001).

\bibitem{john+maciej} J. Schliemann, J.~I. Cirac, M. Ku\'s, M. Lewenstein, and
D. Loss, Phys. Rev. A {\bf 64}, 022303 (2001).

\bibitem{john+kai} K.~Eckert, J.~Schliemann, D.~Bru{\ss}, and M.~Lewenstein,
Annals of Physics {\bf 299}, 1 (2002).

\bibitem{paskauskas} R. Paskauskas and L. You, Phys. Rev. A {\bf 64}, 042310 (2001).

\bibitem{zanardi} P. Zanardi, quant-ph/0104114.

\bibitem{fisher} J. R. Gittings and A. J. Fisher, quant-ph 0202051.

\bibitem{echo}
F. B. J. Buchkremer, R. Dumke, H. Levsen, G. Birkl, and W. Ertmer,
Phys. Rev. Lett. {\bf 85}, 3121 (2000).

\end{thebibliography}
\end{document}